\documentclass{article}%
\usepackage{amsfonts}
\usepackage{amsmath}
\usepackage{amssymb}
\usepackage{graphicx}%
\begin{document}

\title{On the embedding of spacetime in five-dimensional Weyl spaces }
\author{$^{a}$F. Dahia, $^{b}$G. A. T. Gomez, $^{b}$C. Romero\\$^{a}$Departamento de F\'{\i}sica, Universidade Federal de Campina Grande,\\58109-970, Campina Grande, Pb, Brazil\\$^{b}$Departamento de F\'{\i}sica, Universidade Federal da Para\'{\i}ba, \\C. Postal 5008, 58051-970 Jo\~{a}o Pessoa, Pb, Brazil\\e-mail: cromero@fisica.ufpb.br, Brazil}
\maketitle

\begin{abstract}
We revisit Weyl geometry in the context of recent higher-dimensional theories
of spacetime. After introducing the Weyl theory in a modern geometrical
language we present some results that represent extensions of Riemannian
theorems. We consider the theory of \ local embeddings and submanifolds in the
context of Weyl geometries and show how a Riemannian spacetime may be locally
and isometrically embedded in a Weyl bulk. We discuss the problem of classical
confinement and the stability of motion of particles and photons in the
neighbourhood of branes for the case when the Weyl bulk has the geometry of a
warped product space. We show how the confinement and stability properties of
geodesics near the brane may be affected by the Weyl field. We construct a
classical analogue of quantum confinement inspired in theoretical-field models
by considering a Weyl scalar field which depends only on the extra coordinate.

\end{abstract}

\section{Introduction}

It has been suggested in the recent years that our ordinary spacetime may be
viewed as a hypersurface embedded in a higher-dimensional manifold, often
referred to as \textit{the bulk }\cite{Roy}. As far as the geometry of this
hypersurface is concerned, it has been generally assumed that it has a
Riemannian geometrical structure. This assumption avoids possible conflicts
with the well-established theory of general relativity which operates in a
Riemannian geometrical frame. On the other hand, with very few exceptions,
there has not been much discussion\ on what kind of geometry the bulk
possesses, which is generally supposed to be also Riemannian. A\ very few
attempts to broaden this scenario has appeared recently in the literature,
where a non-Riemannian geometry, namely a Weyl geometry, is taken into
consideration as a viable possibility to describe the bulk
\cite{Israelit,Arias,Nandinii}. However, a vast number of interesting
non-Riemannian geometries are currently known and could be investigated in
this context. It is presently our intention to carry out such a program of
research, and a first step in this direction would be to consider the Weyl
geometry \cite{Weyl}, one of the simplest generalizations of Riemannian
geometry. \ \ \ \ \ \ \ 

The paper is organized as follows. We start in Section 2 with a modern
definition of the Weyl geometry and state some results that represent
straightforward extensions of Riemannian theorems. We proceed in Section 3 to
consider the theory of \ local embeddings and submanifolds in the context of
Weyl geometries. In Section 4 we show that a Riemannian spacetime may be
embedded in a Weyl bulk and this constitutes one of our main results. Section
5 contains an application of the formalism to the problem of classical
confinement and the stability of motion of particles and photons in the
neighbourhood of branes. In Section 6 we show how the presence of a Weyl field
may affect both the confinement and/or stablity of the particle's motion and
discuss how a geometrical field, such as the Weyl field, may effectively act
as a quantum scalar field, which in some theoretical-field modes is the
responsible for the confinement of matter in the brane \cite{Rubakov}.
\ \ \ \ \ \ \ \ \ \ \ \ \ \ \ \ \ \ \ \ \ \ \ \ \ \ \ \ \ \ \ \ \ \ \ \ \ \ \ \ \ \ \ \ \ \ \ \ \ \ \ \ \ \ \ \ \ \ \ \ \ \ \ \ \ \ \ \ \ \ \ \ \ \ \ \ \ \ \ \ \ \ \ \ \ \ \ \ \ \ \ \ \ \ \ \ \ \ \ \ \ \ \ \ \ \ \ \ \ \ \ \ \ \ \ \ \ \ \ \ \ \ \ \ \ \ \ \ \ \ \ \ \ \ In
Section 7 we give a simple application of the ideas developed previously. We
conclude, in Section 8, with some final remarks.\ \ \ \ \ \ \ \ \ \ \ \ \ \ \ \ \ \ \ \ \ \ \ \ \ \ \ \ \ \ \ \ \ \ \ \ \ \ \ \ \ \ \ \ \ \ \ \ \ \ \ \ \ \ \ \ \ \ \ \ \ \ \ \ \ \ \ \ \ \ \ \ \ \ \ \ \ \ \ \ \ \ \ \ \ \ \ \ \ \ \ \ \ \ \ \ \ \ \ \ \ \ \ \ \ \ \ \ \ \ \ \ \ \ \ \ \ \ \ \ \ \ \ \ \ \ \ \ \ \ \ \ \ \ \ \ \ \ \ \ \ \ \ \ \ \ \ \ \ \ \ \ \ \ \ \ \ \ \ \ \ \ \ \ \ \ \ \ \ \ \ \ \ \ \ \ \ \ \ \ \ \ \ \ \ \ \ \ \ \ \ \ \ \ \ \ \ \ \ \ \ \ \ \ \ \ \ \ \ \ \ \ \ \ \ \ \ \ \ \ \ 

\section{Weyl geometry}

As is well known, the kind of geometrical structure conceived by H. Weyl in
1918, although admirably ingenious as an attempt to unify gravity with
electromagnetism, turned out to be a failure as a physical theory. In fact,
immediately after have been exposed to Weyl's ideas, Einstein raised strong
objections to the adoption of Weyl geometry in the description of
electromagnetic as well as gravitational phenomena \cite{Pais}. Einstein's
argument was that in a non-integrable geometry it would not be possible the
existence of sharp spectral lines in the presence of an electromagnetic field
since atomic clocks would depend on their past history \cite{Pauli}. It should
be said, however, that a variant of Weyl geometries, known as Weyl integrable
geometry, which does not suffer from the drawback pointed out by Einstein has
attracted the attention of cosmologists some years ago \cite{Novello}.

In this section we review some basic definitions and results that are valid in
Riemannian and Weyl geometry. As we shall see, Weyl geometry may be viewed as
a kind of generalization of Riemannian geometry, and some theorems that will
be presented here are straightforward extensions of corresponding theorems of
the former. However, these extensions have a different and new flavour
especially when they are applied to study geodesic motion. Let us start with
the definition of affine connection .

\textbf{Definition}. Let $M$ be a differentiable manifold and $T(M)$ the set
of all differentiable vector fields on $M$. An \textit{affine connection } is
a mapping $\nabla:T(M)\times T(M)$\ $\rightarrow T(M)$, which is denoted by
$(U,V)\rightarrow\nabla_{U}V$, satisfying the following properties:
\begin{equation}
i)\text{ }\nabla_{fV+gU}W=f\nabla_{V}W+g\nabla_{U}W, \label{affine1}%
\end{equation}%
\begin{equation}
ii)\text{ }\nabla_{V}(U+W)=\nabla_{V}U+\nabla_{V}W, \label{affine2}%
\end{equation}%
\begin{equation}
iii)\text{ }\nabla_{V}(fU)=V[f]U+f\nabla_{V}U, \label{affine3}%
\end{equation}
where $V,$ $U,$ $W\in T(M),$ $f$ and $g$ are \ $C^{\infty}$\ scalar functions
defined on $M.$ An important result comes immediately from the above
definition and allows one to define a covariant derivative along a
differentiable curve.

\textbf{Proposition}. \textbf{ }Let $M$ be a differentiable manifold endowed
with an affine connection $\nabla$, $V$ a vector field defined along a
differentiable curve $\alpha:(a,b)\subset R\rightarrow M$. Then, there exists
a unique rule which associates another vector field $\frac{DV}{d\lambda}$
along $\alpha$ to $V$, such that
\begin{equation}
i)\text{ }\frac{D(V+U)}{d\lambda}=\frac{DV}{d\lambda}+\frac{DU}{d\lambda},
\label{covariant1}%
\end{equation}%
\begin{equation}
ii)\text{ }\frac{D(fV)}{d\lambda}=\frac{df}{d\lambda}V+f\frac{DV}{d\lambda},
\label{covariant2}%
\end{equation}
where $\alpha=$ $\alpha(\lambda)$ and $\lambda$ $\in(a,b)$.

$\qquad\qquad iii)$ If the vector field $U(\lambda)$ is induced by a vector
field $\hat{U}$ $\in T(M)$, then $\frac{DU}{d\lambda}=\nabla_{V}$ $U$, where
$V$ is the tangent vector to the curve $\alpha$, i.e $V=\frac{d}{d\lambda}.$
For a proof of this proposition we refer the reader to \cite{do Carmo}. We now
are ready for being introduced to the concept of parallel transport of a
vector\ along a given curve.

\textbf{Definition}. Let $M$ be a differentiable manifold \ with an affine
connection $\nabla$, $\alpha:(a,b)\in R\rightarrow M$ a differentiable curve
on $M$, and $V$ a vector field defined along $\alpha=$ $\alpha(\lambda)$.\ The
vector field $V$ is said to be \textit{parallel }if $\frac{DV}{d\lambda}=0$
for any value of the parameter $\lambda\in(a,b)$.

Among all admissible affine connections defined on a manifold, an important
role in Riemannian and also in Weyl theory is played by a special class of
connections, namely, the\textit{ torsionless }connections, as defined below.

\textbf{Definition}. \textbf{ }We say that an affine connection $\nabla$
defined on $M$ is \textit{torsionless} \ (or \textit{symmetric}) if for any
$U,V\in T(M)$ the following condition holds:
\begin{equation}
\nabla_{V}U-\nabla_{U}V=[V,U] \label{torsionless}%
\end{equation}

We now introduce the concept of Weyl manifold through the following
definition. \qquad\qquad\ \ 

\textbf{Definition}. Let $M$ be a differentiable manifold \ endowed with an
affine connection $\nabla$, a metric tensor $g$ and a one-form field $\sigma$,
called a Weyl field, globally defined in $\ M$. We say that $\nabla$ is
Weyl-compatible ( or W-compatible) with $g$ if for any vector fields $U,V,$
$W\in T(M)$,\ the following condition is satisfied:%

\begin{equation}
V[g(U,W)]=g(\nabla_{V}U,W)+g(U,\nabla_{V}W)+\sigma(V)g(U,W)
\label{W-compatible}%
\end{equation}
This is, of course, a generalization of the idea of Riemannian compatibility
between $\nabla$ and $g$. If the one-form $\sigma$ vanishes throughout $M$,
\ we recover the Riemannian compatibility condition. It is therefore rather
natural to expect that a generalized version of the Levi-Civita theorem holds
if we restrict ourselves to torsionless connections. Indeed, we have the
following result:

\textbf{Theorem (Levi-Civita extended)}. In a given differentiable manifold
$M$ endowed with a metric $g$ and a differentiable one-form\ field $\sigma$
defined on $M$, there exists only one affine connection $\nabla$ such that:
$i)$ $\nabla$ is torsionless; $ii)$ $\nabla$ is W-compatible.

Proof. Let us first suppose that such $\nabla$ exists. Then, from
(\ref{W-compatible}) we have the following three equations%
\begin{equation}
V[g(U,W)]=g(\nabla_{V}U,W)+g(U,\nabla_{V}W)+\sigma(V)g(U,W) \label{1}%
\end{equation}%
\begin{equation}
W[g(V,U)]=g(\nabla_{W}V,U)+g(V,\nabla_{W}U)+\sigma(W)g(V,U) \label{2}%
\end{equation}%
\begin{equation}
U[g(W,V)]=g(\nabla_{U}W,V)+g(W,\nabla_{U}V)+\sigma(U)g(W,V) \label{3}%
\end{equation}
Adding (\ref{1}) and (\ref{2}) and subtracting (\ref{3}), and also taking into
account the torsionless condition (\ref{torsionless}), we are left with
\begin{align}
g(\nabla_{W}V,U)  &  =\frac{1}{2}%
V[g(U,W)]+W[g(V,U)]-U[g(W,V)]-g([V,W],U)\label{W}\\
&  -g([W,U],V)-g([V,U],W)+\sigma(U)g(V,W)-\sigma(V)g(U,W)-\sigma
(W)g(U,V)\}\nonumber
\end{align}
The above equation shows that the affine connection $\nabla$, if it exists, is
uniquely determined from the metric $g$ and the Weyl field of one-forms
$\sigma$. Now, to prove the existence of such a connection we just define
$\nabla_{U}V$ by means of (\ref{W}). At this point it is instructive to write
(\ref{W-compatible}) in a local coordinate system $\left\{  x^{a}\right\}  $,
$a=1,...,n$. Then, a straightforward calculation shows that one can express
the components of the affine connection completely in terms of the components
of $g$ and $\sigma$:%
\begin{equation}
\Gamma_{bc}^{a}=\{_{bc}^{a}\}-\frac{1}{2}g^{ad}[g_{db}\sigma_{c}+g_{dc}%
\sigma_{b}-g_{bc}\sigma_{d}] \label{Weylconnection}%
\end{equation}
where $\{_{bc}^{a}\}=$ $\frac{1}{2}g^{ad}[g_{db,c}+g_{dc,b}-g_{bc,d}]$ denotes
the Christoffel symbols of second kind.

At this stage let us note that the Weyl compatibility condition
(\ref{W-compatible})\ may equivalently be interpreted as requiring that the
covariant derivative of the metric tensor $g$ in the direction of a vector
field $V\in T(M)$ do not vanish, as in Riemannian geometry, but, instead, that
it be regulated by the Weyl field $\sigma$ defined in the manifold $M$. In
\ other words, we must have
\begin{equation}
\nabla g=\sigma\otimes g \label{covariant}%
\end{equation}
where $\sigma\otimes g$ denotes the direct product of $\sigma$ and $g$. \ It
looks a bit surprising that this new requirement does not spoil the miraculous
determinability of the connection $\nabla$ from $\sigma$ and $g$ only.
\cite{Berger}. A clear geometrical insight on the properties of Weyl parallel
transport is given by the following proposition:

\textbf{Corollary }Let $M$ be a differentiable manifold with an affine
connection $\nabla$, a metric $g$ and a Weyl field of one-forms $\sigma$. If
$\nabla$ is W-compatible, then for any smooth curve $\alpha=\alpha(\lambda)$
and any pair of two parallel vector fields $V$ and $U$ along $\alpha,$ we
have
\begin{equation}
\frac{d}{d\lambda}g(V,U)=\sigma(\frac{d}{d\lambda})g(V,U)
\label{covariantderivative}%
\end{equation}
where $\frac{d}{d\lambda}$ denotes the vector tangent to $\alpha$.

If we integrate the above equation along the curve $\alpha$, starting from a
point $P_{0}=\alpha(\lambda_{0})$ then we readily obtain%
\begin{equation}
g(V(\lambda),U(\lambda))=g(V(\lambda_{0}),U(\lambda_{0}))e^{\int_{\lambda_{0}%
}^{\lambda}\sigma(\frac{d}{d\rho})d\rho} \label{integral}%
\end{equation}
Putting $U=V$ and denoting by $L(\lambda)$ the length of the vector
$V(\lambda)$ at an arbitrary point\ $P=\alpha(\lambda)$\ of the curve, then it
is easy to see that in a local coordinate system $\left\{  x^{a}\right\}  $
the equation (\ref{covariantderivative}) reduces to
\[
\frac{dL}{d\lambda}=\frac{\sigma_{a}}{2}\frac{dx^{a}}{d\lambda}L
\]

Consider the set of all closed curves $\alpha:[a,b]\in R\rightarrow M$, i.e,
with $\alpha(a)=\alpha(b).$ Then, the equation
\[
g(V(b),U(b))=g(V(a),U(a))e^{\int_{a}^{b}\sigma(\frac{d}{d\lambda})d\lambda}%
\]
defines a holonomy group, whose elements are, in general, a composition of a
homothetic transformation and an isometry. If we want the elements of this
group to correspond to an isometry only, then we must have%

\[%
{\displaystyle\oint}
\sigma(\frac{d}{d\lambda})d\lambda=0
\]
for any loop. It follows, from Stokes' theorem, that $\sigma$ must be an exact
form, that is, there exists a scalar function $\phi$, such that $\sigma=d\phi
$. In this case we have what is often called in the literature a \textit{Weyl
integrable manifold}.

Weyl manifolds are completely caracterized by the triple $(M,g,\sigma)$, which
we shall call a \textit{Weyl frame}.\ It is interesting to see that the Weyl
compatibility condition (\ref{covariantderivative}) remains unchanged\ when we
go to another Weyl frame $(M,\overline{g},\overline{\sigma})$ by performing
the following simultaneous transformations in $g$ and $\sigma$:%
\begin{equation}
\overline{g}=e^{-\phi}g \label{conformal}%
\end{equation}%
\begin{equation}
\overline{\sigma}=\sigma-d\phi\label{gauge}%
\end{equation}
where $\phi$ is a scalar function defined on $M$. \ Clearly the conformal map
(\ref{conformal}) and the \textit{gauge }transformation (\ref{gauge}) \ define
classes of equivalences in the set of Weyl frames. It is worth mentioning that
the discovery that the compatibility requirement (\ref{covariantderivative})
is invariant under this group of transformations was what primarely led Weyl
to his attempt at unifying gravity and electromagnetism, extending the concept
of spacetime to that of a collection of manifolds equipped with a conformal
structure, i.e, the spacetime would be viewed as a class $[g]$ of conformally
equivalent Lorentzian metrics \cite{Weyl}.

\bigskip

\section{Submanifolds and isometric embeddings in Weyl geometry}

\textbf{Definition}. \textbf{ } Let $(M,g,\sigma)$ \ and $(\overline{M}%
$,$\overline{g},\overline{\sigma})$ be differentiable Weyl manifolds of
dimensions $m$ and $n=m+k$, respectively. A differentiable map $f:M\rightarrow
\overline{M\text{ }}$ is called an \textit{immersion }if

$i)$\ the\ differential $f_{\ast}:T_{P}(M)$ $\rightarrow T_{f(P)}\overline{M}$
is injective for any $P\in M$;

$ii)$ $\sigma(V)=\overline{\sigma}($ $f_{\ast}(V))$ for any $V\in T_{P}(M)$.
The number $k$ is called the \textit{codimension }of $f$. We say that the
immersion $f:M\rightarrow\overline{M\text{ }}$ is \textit{isometric} at a
point $P\in M$ if $g(U,V)=\overline{g}($ $f_{\ast}(U),$ $f_{\ast}(V))$ for
every $U,V$ in the tangent space $T_{P}(M)$. If, in addition, $f$ is a
homeomorphism onto $f(M)$, then we say that $f$ is an \textit{embedding.}
\bigskip If $M\subset\overline{M}$ and the inclusion $i:M\subset\overline{M}$
$\rightarrow\overline{M}$ is an embedding, then $M$ is\ called a
\textit{submanifold} of $\overline{M}$.

It is important to note that locally any immersion is an embedding. Indeed,
let $f:M\rightarrow\overline{M\text{ }}$ be an immersion. Then, around each
$P\in M$, there is a neighbourhood $U\subset M$ such that the restriction of
$f$ to $U$ is an embedding onto $f(U).$ We may, therefore, identify $U$ with
its image under $f$, so that locally we can regard $M$ as a submanifold
embedded in $\overline{M}$, with $f$ \ actually being the inclusion map. Thus,
we shall identify each vector $V\in T_{P}(M)$ with $f_{\ast}(V)\in
T_{f(P)}(\overline{M})$ and consider $T_{P}(M)$ as a subspace of
$T_{f(P)}(\overline{M}).$

Now, in the vector space $T_{P}(\overline{M})$ the metric $\overline{g}$
allows one to make the decomposition $T_{P}(\overline{M})=T_{P}(M)\oplus
T_{P}(M)^{\bot},$ where $T_{P}(M)^{\bot}$ is the orthogonal complement of
$T_{P}(M)$ $\subset T_{P}(\overline{M})$. That is, for any vector
$\overline{V}\in T_{P}(\overline{M})$, with $P\in M$, we can decompose
$\overline{V}$ into $\overline{V}=V+V^{\bot},$ $V\in T_{P}(M)$, \ $V^{\bot}\in
T_{P}(M)^{\bot}$.

Let us denote the Weyl connection on $\overline{M}$ by $\overline{\nabla}.$We
now can prove the following proposition.

\textbf{Proposition}. \textit{If }$V$\textit{ and }$U$\textit{ are local
vector fields on }$M$\textit{, and }$\overline{V}$\textit{ and }$\overline{U}%
$\textit{ are local extensions of these fields to }$\overline{M}$\textit{,
then the Weyl connection }$\nabla_{V}U$\textit{ will be given by }%
\begin{equation}
\nabla_{V}U=(\overline{\nabla}_{\overline{V}}\overline{U})^{\top} \label{c}%
\end{equation}
\textit{where }$(\overline{\nabla}_{\overline{V}}\overline{U})^{\top}$\textit{
is the tangential component of }$\overline{\nabla}_{\overline{V}}\overline{U}%
$\textit{. }

Proof. We start with the equation which expresses the Weyl compatibility
requirement%
\begin{equation}
\overline{V}[\overline{g}(\overline{U},\overline{W})]=\overline{g}%
(\overline{\nabla}_{\overline{V}}\overline{U},\overline{W})+\overline
{g}(\overline{U},\overline{\nabla}_{\overline{V}}\overline{W})+\overline
{\sigma}(\overline{V})\overline{g}(\overline{U},\overline{W}) \label{W1}%
\end{equation}
where $\overline{V}$, $\overline{U}$ , $\overline{W}$ \ $\in$ $T(\overline
{M}).$ Now, suppose that $\overline{V}$, $\overline{U}$ , $\overline{W}$ are
local extensions of the the vector fields $V,U,W$ to $\overline{M}.$ Clearly,
at a point $P\in M$, we have
\begin{equation}
\overline{V}[\overline{g}(\overline{U},\overline{W})]=V[\overline{g}%
(\overline{U},\overline{W})]=V[g(U,W)] \label{a}%
\end{equation}
where we have taking into account that the inclusion of $M$ into
$\overline{M\text{ }}$ is isometric. On the other hand, evaluating separately
each term of the right-hand side of (\ref{W1}) at $\ P$ yields%
\begin{equation}
\overline{g}(\overline{\nabla}_{\overline{V}}\overline{U},\overline
{W})=\overline{g}((\overline{\nabla}_{\overline{V}}\overline{U})^{\top
}+(\overline{\nabla}_{\overline{V}}\overline{U})^{\bot},\overline
{W})=\overline{g}((\overline{\nabla}_{\overline{V}}\overline{U})^{\top
},\overline{W})=g(\overline{\nabla}_{\overline{V}}\overline{U})^{\top},W)
\label{b}%
\end{equation}
with an analogous expression for $\overline{g}(\overline{U},\overline{\nabla
}_{\overline{V}}\overline{W})$. From the above equations and the fact that
$\overline{\sigma}(\overline{V})=\sigma(V)$ \ we finally obtain
\[
V[g(U,W)]=g((\overline{\nabla}_{\overline{V}}\overline{U})^{\top
},W)+g(U,(\overline{\nabla}_{\overline{V}}\overline{W})^{\top})+\sigma
(V)g(U,W)
\]
From the Levi-Civita theorem extended to Weyl manifolds, which asserts the
uniqueness of affine connection $\nabla$ in a Weyl manifold we conclude that
(\ref{c}) \ holds. In other words, the tangential component of the covariant
derivative $\overline{\nabla}_{\overline{V}}\overline{U}$, \ evaluated at
points of $M$, is nothing more than the covariant derivative of the induced
Weyl connection from the metric $g$ on $M$, defined by $g(V,U)=\overline{g}($
$f_{\ast}(V),$ $f_{\ast}(U)).$

\bigskip

\section{Embedding the spacetime in a Weyl bulk}

Now that we know how the mechanism of embedding submanifolds in Weyl geometry
works, we are led to ask the following question: Is it possible to have a
Riemannian submanifold embedded in a Weyl ambient space? The answer to this
question is given by the following argument. A Riemannian manifold is a
particular case of a Weyl manifold, in which the Weyl field $\sigma$ vanishes.
Therefore, a submanifold $M$ embedded in Weyl space $\overline{M}$ will be
Riemannian if and only if the field of 1-forms $\sigma$ induced by pullback
from $\overline{\sigma}$ vanishes throughout $M$. That is, the necessary and
sufficient condition for $M$ to be an embedded Riemannian manifold is that
$\sigma(V)=0$ for any $V\in T(M)$.

To illustrate the above, and having in view future applications, let us
consider the case in which the manifold $\overline{M}$ is foliated by a family
of submanifolds defined by $k$ equations $y^{A}=y_{o}^{A}=$constant
\footnote{From now on lower case Latin indices take value in the range
$(0,1,...,(n+3))$, while Greek indices run over $(0,1,2,3).$ The coordinates
of a generic point $P$ of the manifold $\overline{M}$ will be denoted by
$y^{a}=(x^{\alpha},y^{4},...y^{n+3})$, where $x^{\alpha}$ denotes the
four-dimensional spacetime coordinates and $y^{A}(A>3)$ refers to the $n$
extra coordinates of $P$.}, with the spacetime $M$ corresponding to one of
these manifolds $y^{A}=y_{o}^{A}=$constant. In local coordinates $\left\{
y^{a}\right\}  $ of $\overline{M}$\ adapted to the embedding the condition
$\sigma(V)=0$ reads $\sigma_{\alpha}V^{\alpha}=0$, where $\sigma=$ $\sigma
_{a}dx^{a}$ and $V=V^{\beta}\partial_{\beta}$. In the case of a Weyl
integrable manifold $\sigma=d\phi$. In this case $\sigma(V)=0$ \ for any $V\in
T(M)$ if, and only if $\frac{\partial\phi}{\partial x^{\alpha}}=0.$ Therefore,
in a Weyl integrable manifold\ if the scalar field $\phi$ is a function of the
extra coordinates only, then the spacetime submanifold $M$ embedded in the
bulk $\overline{M}$ is Riemannian.

The fact that we may have a Riemannian spacetime $M$ embedded in a Weyl bulk
$\overline{M}$ does not mean that physical or geometrical effects coming from
the extra dimensions should be absent. A nice illustration of this point is
given by the behaviour of geodesics near the $M$. In section 6 we shall
examine how a Weyl field may affect the geodesic motion in the case of a bulk
with a warped product geometry. We shall be interested particularly in the
problem of confinement and stability of the motion of particles and photons
near the spacetime submanifold. \cite{Seahra, Dahia}

\section{Geodesic motion in a Riemannian warped product space}

In this section let us consider the case where the geometry of the bulk
contains two special ingredients: a) It is a Riemannian manifold and \ b) its
metric has the structure of a warped product space \cite{Frolov}. As is well
known, the importance of warped product geometry is closely related to the
so-called braneworld scenario \cite{Roy}. Let us start with the investigation
of geodesics in warped product spaces, firstly considering the Riemannian case.

We define a warped product space in the following way. Let $(M,g)$ and $(N,h)$
be two\ Riemannian manifolds of dimension $m$ and $r$, with metrics $g$ and
$h,$ respectively. Suppose we are given a smooth function $f:N\rightarrow R$
(which will\textbf{ }called \textit{warping function}). Then we can construct
a new Riemannian manifold by setting $\overline{M}=M\times N$ and defining a
metric $\overline{g}=e^{2f}g\oplus k$. Here, for simplicity, we shall take
$M=M^{4}\mathbb{\ }$and $N=R$, where $M^{4}$ denotes a four-dimensional
Lorentzian manifold with signature $(+---)$ (referred to as \textit{spacetime}%
). In local coordinates $\{y^{a}=(x^{\alpha},y^{4}\}$ the line element
corresponding to this metric will be written as \footnote{Throughout this
section Latin indices take values in the range (0,1,...4) while Greek indices
run from (0,1,2,3).}
\[
dS^{2}=\overline{g}_{ab}dy^{a}dy^{b}%
\]

The equations of geodesics in the five-dimensional space $\overline{M}$ will
be given by%
\begin{equation}
\frac{d^{2}y^{a}}{d\lambda^{2}}+^{(5)}\Gamma_{bc}^{a}\frac{dy^{b}}{d\lambda
}\frac{dy^{c}}{d\lambda}=0, \label{geodesics5D}%
\end{equation}
where $\lambda$ is an affine parameter and $^{(5)}\Gamma_{bc}^{a}$ denotes the
5D Christoffel symbols of the second kind defined by $^{(5)}\Gamma_{bc}%
^{a}=\frac{1}{2}\overline{g}^{ad}\left(  \overline{g}_{db,c}+\overline
{g}_{dc,b}-\overline{g}_{bc,d}\right)  $. Denoting the fifth coordinate
$y^{4}$ by $y$ and the remaining coordinates $y^{\mu}$\ (the "spacetime"
\textit{\ }coordinates) by $x^{\mu}$, i.e. $y^{a}=(x^{\mu},y)$, we can easily
show that the "4D part" of the geodesic equations (\ref{geodesics5D}) can be
rewritten in the form
\begin{equation}
\frac{d^{2}x^{\mu}}{d\lambda^{2}}+^{(4)}\Gamma_{\alpha\beta}^{\mu}%
\frac{dx^{\alpha}}{d\lambda}\frac{dx^{\beta}}{d\lambda}=\xi^{\mu},
\label{4Dpart}%
\end{equation}
where
\begin{align}
\xi^{\mu}  &  =-^{(5)}\Gamma_{44}^{\mu}\left(  \frac{dy}{d\lambda}\right)
^{2}-2^{(5)}\Gamma_{\alpha4}^{\mu}\frac{dx^{\alpha}}{d\lambda}\frac
{dy}{d\lambda}\nonumber\\
&  -\frac{1}{2}\overline{g}^{\mu4}\left(  \overline{g}_{4\alpha,\beta
}+\overline{g}_{4\beta,\alpha}-\overline{g}_{\alpha\beta,4}\right)
\frac{dx^{\alpha}}{d\lambda}\frac{dx^{\beta}}{d\lambda},
\end{align}
and $^{(4)}\Gamma_{\alpha\beta}^{\mu}=\frac{1}{2}\overline{g}^{\mu\nu}\left(
\overline{g}_{\nu\alpha,\beta}+\overline{g}_{\nu\beta,\alpha}-\overline
{g}_{\alpha\beta,\nu}\right)  $.

At this point we turn our attention to the five-dimensional brane-world
scenario, where the bulk corresponds to the five-dimensional manifold
$\overline{M}$, which, as in the previous section, is assumed to be foliated
by a family of submanifolds (in this case, hypersurfaces) defined by the
equation $y=$ constant.

It turns out that the geometry of a generic hypersurface, say \ $y=y_{0},$
will be determined by the induced metric $g_{\alpha\beta}(x)=\overline
{g}_{\alpha\beta}(x,y_{0})$. Thus, on the hypersurface we have%
\[
ds^{2}=\overline{g}_{\alpha\beta}(x,y_{0})dx^{\alpha}dx^{\beta}.
\]
We see then that the quantities $^{(4)}\Gamma_{\alpha\beta}^{\mu}$ which
appear on the left-hand side of Eq. (\ref{4Dpart}) are to be identified with
the Christoffel symbols associated with the induced metric in the leaves of
the foliation defined above.

Let us now consider the class of warped geometries given by the following line
element
\begin{equation}
dS^{2}=e^{2f}g_{\alpha\beta}dx^{\alpha}dx^{\beta}-dy^{2}, \label{warped}%
\end{equation}
where $f=f(y)$ and $g_{\alpha\beta}=$ $g_{\alpha\beta}(x)$. For this metric it
is easy to see\footnote{In the above calculation we have used the fact that
the matrix $g_{\alpha\beta}$ has an inverse $g^{\alpha\beta}$, that is,
$\ g^{\mu\beta}g_{\beta\nu}=\delta_{\nu}^{\mu}$. This may be easily seen since
by definition $\det g=-\det\overline{g}\neq0$.} that $^{(5)}\Gamma_{44}^{\mu
}=0$ and $^{(5)}\Gamma_{4\nu}^{\mu}=\frac{1}{2}\overline{g}^{\mu\beta
}\overline{g}_{\beta\nu,4}=f^{\prime}\delta_{\nu}^{\mu}$, where a prime
denotes a derivative with respect to $y$. Thus in the case of the warped
product space (\ref{warped}) the right-hand side of Eq. (\ref{4Dpart}) reduces
to $\xi^{\mu}=-2f^{\prime}\frac{dx^{\mu}}{d\lambda}\frac{dy}{d\lambda}$ and
the 4D part of the geodesic equations becomes
\begin{equation}
\frac{d^{2}x^{\mu}}{d\lambda^{2}}+^{(4)}\Gamma_{\alpha\beta}^{\mu}%
\frac{dx^{\alpha}}{d\lambda}\frac{dx^{\beta}}{d\lambda}=-2f^{\prime}%
\frac{dx^{\mu}}{d\lambda}\frac{dy}{d\lambda}. \label{4Dwarped}%
\end{equation}
On the other hand the geodesic equation for the fifth coordinate $y$ in the
warped product space becomes%

\begin{equation}
\frac{d^{2}y}{d\lambda^{2}}+f^{\prime}e^{2f}g_{\alpha\beta}\frac{dx^{\alpha}%
}{d\lambda}\frac{dx^{\beta}}{d\lambda}=0. \label{lwarped}%
\end{equation}
By restricting ourselves to 5D timelike geodesics $\left(  \overline{g}%
_{ab}\frac{dy^{a}}{d\lambda}\frac{dy^{b}}{d\lambda}=1\right)  $ we can readily
decouple the above equation from the 4D spacetime coordinates to obtain
\begin{equation}
\frac{d^{2}y}{d\lambda^{2}}+f^{\prime}\left(  1+\left(  \frac{dy}{d\lambda
}\right)  ^{2}\right)  =0. \label{5Dmotion}%
\end{equation}
Similarly, to study the motion of photons in 5D, we must consider the null
geodesics $\left(  \overline{g}_{ab}\frac{dy^{a}}{d\lambda}\frac{dy^{b}%
}{d\lambda}=0\right)  $, in which case Eq. (\ref{lwarped}) becomes
\begin{equation}
\frac{d^{2}y}{d\lambda^{2}}+f^{\prime}\left(  \frac{dy}{d\lambda}\right)
^{2}=0. \label{5Dphoton}%
\end{equation}

\bigskip Equations (\ref{5Dmotion}) and (\ref{5Dphoton}) are ordinary
differential equations of second-order which, in principle, can be solved if
the function $f^{\prime}=f^{\prime}(y)$ is known. A qualitative picture of the
motion in the fifth dimension may be obtained without the need to solve
(\ref{5Dmotion}) and (\ref{5Dphoton}) analytically \cite{Dahia}. This is done
by defining the variable $q=\frac{dy}{d\lambda}$ and then investigating the
autonomous dynamical system \cite{Andronov}
\begin{align}
\frac{dy}{d\lambda}  &  =q\nonumber\label{dynamical}\\
\frac{dq}{d\lambda}  &  =P(q,y)
\end{align}
with $P(q,y)=-f^{\prime}(\epsilon+q^{2})$, where$\ \epsilon=1$ in the case of
(\ref{5Dmotion}) (corresponding to the motion of particles with nonzero rest
mass) and $\epsilon=0$ in the case of (\ref{5Dphoton}) (corresponding to the
motion of photons). In the investigation of dynamical systems a crucial role
is played by their \textit{equilibrium points}, which in the case of system
(\ref{dynamical}) are given by $\frac{dy}{d\lambda}=0$ and $\frac{dq}%
{d\lambda}=0$. The knowledge of these points together with their stability
properties provides a great deal of information on the types of behaviour
allowed by the system.

\subsection{The case of massive particles}

In the case of nonzero rest mass particles, the motion in the fifth dimension
is governed by the dynamical system
\begin{align}
\frac{dy}{d\lambda}  &  =q\label{dynamical1}\\
\frac{dq}{d\lambda}  &  =-f^{\prime}(1+q^{2})
\end{align}
The critical points of (\ref{dynamical1}) are given by $q=0$ and the zeros of
the function $f^{\prime}(y)$ (if they exist) which we generically denote by
$y_{0}$. These solutions, pictured as isolated points in the phase plane,
correspond to curves which lie entirely on a hypersurface $M$ of our foliation
(since for them $y=$ constant). It turns out that these curves are timelike
geodesics with respect to the hypersurface induced geometry \cite{Dahia}.

To obtain information about the possible modes of behaviour of particles and
light rays in such hypersurfaces, it is important to study the nature and
stability of the corresponding equilibrium points. This can be done by
linearising equations (\ref{dynamical1}) and studying the eigenvalues of the
corresponding Jacobian matrix about the equilibrium points. Assuming that the
function $f^{\prime}(y)$ vanishes, at least at one point $y_{0}$, it can
readily be shown that the corresponding eigenvalues are determined by the sign
of the second derivative $f^{\prime\prime}(y_{0})$, at the equilibrium point,
and some possibilities arise for the equilibrium points of the dynamical
system (\ref{dynamical1}) \cite{Dahia}. We shall discuss only \ the three
following cases.

Case I. If $f^{\prime\prime}(y_{0})>0$, then the equilibrium point
$(q=0,y=y_{0})$ is a \textit{center}. This corresponds to the case in which
the massive particles oscillate about the hypersurface $M$ $(y=y_{0}).$ Such
cyclic motions are independent of the ordinary 4D spacetime dimensions, and,
except for the conditions $f^{\prime}(y_{0})=0$ and\ $f^{\prime\prime}%
(y_{0})>0$, the warping function $f(y)$ remains completely arbitrary.

Case II. If $f^{\prime\prime}(y_{0})<0$, then the point $(q=0,y=y_{0})$ is a
\textit{saddle point}. In this case the solution corresponding to the
equilibrium point is highly unstable and the smallest transversal perturbation
in the motion of particles along the brane will cause them to be expelled into
the extra dimension. An example of this highly unstable "confinement" \ at the
hypersurface $y=0$ is provided by Gremm's warping function \cite{Gremm}\
\begin{equation}
f(y)=-b\ln\cosh(cy), \label{warpfunction}%
\end{equation}
where $b$ and $c$ are positive constants.

Case III. There are no equilibrium points at all. The warping function $f(y)$
does not have any turning points for any value of $y$. This implies that in
this case we cannot have confinement of classical particles to hypersurfaces
solely due to gravitational effects. An example of this situation is
illustrated by the warping function $f(y)=\frac{1}{2}\ln\left(  \Lambda
y^{2}/3\right)  $ considered in \cite{Mashhoon}. In similar fashion, note that
for large values of $y$ the warping function (\ref{warpfunction}) approaches
that of the Randall-Sundrum metric \cite{Randall}) \newline%
\[
ds^{2}=e^{-2k\left\vert y\right\vert }\eta_{\alpha\beta}dx^{\alpha}dx^{\beta
}-dy^{2},
\]
where $k$ is a constant. In this case $f^{\prime}(y)=\mp k$ according to
whether $y$ is positive or negative. Again, there exist no equilibrium points,
and therefore no confinement of particles is possible due only to gravity.

\subsection{The case of photons}

The motion of photons is governed by the dynamical system
\begin{align}
\frac{dy}{d\lambda}  &  =q\nonumber\label{dynamical2}\\
\frac{dq}{d\lambda}  &  =-f^{\prime}q^{2}.
\end{align}
The equilibrium points in this case are given by $q=0$, so they consist of a
line of equilibrium points along the $y$-axis, with eigenvalues both equal to
zero. Any point along the $y$-axis is an equilibrium point and corresponds to
a 5D null geodesics in the hypersurface $y=$ constant. The existence of
photons confined to hypersurfaces does not depend upon the warping factor
\cite{Dahia}.

As is well known, in the brane-world scenario the stability of the confinement
of matter fields at the quantum level is made possible by assuming an
interaction of matter with a scalar field. An example of how this mechanism
works is clearly illustrated \ by a field-theoretical model devised by
Rubakov, in which fermions may be trapped to a brane by interacting with a
scalar field that depends only on the extra dimension \cite{Rubakov}. On the
other hand, the kind of confinement we are concerned with is purely
geometrical, and that means the only force acting on the particles is the
gravitational force. In a purely classical (non-quantum) picture, one would
like to have effective mechanisms other than a quantum scalar field in order
to constrain massive particles to move on hypersurfaces in a stable way. At
this point at least two possibilities come to our mind. One is to assume a
direct interaction between the particles and a physical scalar field.
Following this approach it has been shown that stable confinement in a thick
brane is possible by means of a direct interaction of the particles with a
scalar field through a modification of the Lagrangian of the particle
\cite{Dahia1}. Another approach would appeal to pure geometry: for instance,
modelling the bulk with a Weyl geometrical structure. As we shall see, in this
case the Weyl field may provide the mechanism necessary for confinement and
stabilization of the motion of particles in the brane.

\section{Geodesic motion in the presence of a Weyl field}

The question we want to discuss now is: What happens with the geodesic motion
pictured in the previous section when we "turn on" a Weyl field? For
simplicity, let us consider the case when the warped product bulk is an
integrable Weyl manifold $(\overline{M},\overline{g,}\phi)$. As we have seen
in Section IV, if the Weyl scalar depends only on the extracoordinates, then
the Weyl field of 1-forms $\sigma=d\phi$ induced on the hypersurfaces of the
foliation defined above vanishes. Indeed, any tangent vector $V$ of a given
leaf $M$ has the form $V=V^{\alpha}\partial_{\alpha}.$ Thus, we have
$\sigma(V)=d\phi(V)=V^{\alpha}\frac{\partial\phi}{\partial x^{\alpha}}=0$.
\ Therefore, if $M$ represents our spacetime embedded in a integrable Weyl
bulk $\overline{M}$ \ with $\phi=\phi(y)$, then we can be sure that $M$ has a
Riemannian structure.

We have seen in Section II, that in a Weyl manifold the coefficients of the
Weyl connection $\Gamma_{bc}^{a}$\ are related to the Christoffel symbols
through the equation
\begin{equation}
\Gamma_{bc}^{a}=\{_{bc}^{a}\}-\frac{1}{2}g^{ad}[g_{db}\sigma_{c}+g_{dc}%
\sigma_{b}-g_{bc}\sigma_{d}]\
\end{equation}
From (\ref{Weylconnection}) it is not difficult to show that the geodesic
equation for the fifth coordinate $y$ in this warped product space leads, for
massive particles, to the equation
\begin{equation}
\frac{d^{2}y}{d\lambda^{2}}+f^{\prime}\left(  1+\left(  \frac{dy}{d\lambda
}\right)  ^{2}\right)  -\phi^{\prime}\left(  \frac{1}{2}+\left(  \frac
{dy}{d\lambda}\right)  ^{2}\right)  =0. \label{dynWeyl}%
\end{equation}
where $\phi^{\prime}=\frac{d\phi}{dy}.$

On the other hand, for photons we now have%
\begin{equation}
\frac{d^{2}y}{d\lambda^{2}}+(f^{\prime}-\phi^{\prime})\left(  \frac
{dy}{d\lambda}\right)  ^{2}=0. \label{dynWeyl2}%
\end{equation}

The equations (\ref{dynWeyl}) and (\ref{dynWeyl2}) respectively define the
following dynamical systems:%
\begin{align}
\frac{dy}{d\lambda}  &  =q\\
\frac{dq}{d\lambda}  &  =(\phi^{\prime}-f^{\prime})q^{2}+\frac{\phi^{\prime}%
}{2}-f^{\prime}%
\end{align}

\begin{align}
\frac{dy}{d\lambda}  &  =q\\
\frac{dq}{d\lambda}  &  =(\phi^{\prime}-f^{\prime})q^{2}%
\end{align}
\qquad

Clearly, the presence of the derivative of the Weyl scalar in the above
equations may completely change the picture of the solutions determined by the
dynamical system considered in the previous section. This is because the
existence of equilibrium points, their topology and stability properties now
depends not only on the values the derivatives the warping function take at
the brane, but also on the derivatives of the Weyl scalar field $\phi(y).$

Finally, note that in the case of photons the Weyl scalar field $\phi$ has no
influence on the confinement. This can be easily explained by the fact that,
according to (\ref{conformal }) and (\ref{gauge}), the presence of a scalar
Weyl is equivalent to perform a conformal transformation in the Riemannian
metric $\overline{g}=e^{2f}g\oplus k$. This essentially results in changing
the warping function from $f$ to $f-\phi/2$. Because the\ existence of
confined photons in the hypersurface is independent of the warping function
\cite{Dahia}, the Weyl scalar has no effect in the confinement. This
interesting property can also be explained by the fact that a conformal
transformation does not alter the light-cone structure of a manifold.

\section{A simple example}

As an illustration of the results obtained in the previous section, let us
consider the five-dimensional Riemannian space $\overline{M}$ endowed with a
Mashhoon-Wesson-type metric \cite{Mashhoon}%
\begin{equation}
dS^{2}=\frac{\Lambda^{2}}{3}y^{2}g_{\alpha\beta}dx^{\alpha}dx^{\beta}-dy^{2}.
\label{Mashhoon}%
\end{equation}
As we have remarked in Section 5, in this case \ there is no confinement of
particles in the hypersurfaces $y=const.$ Now let us "turn on" a Weyl field in
the space $\overline{M}$ by chosing, for instance,
\begin{equation}
\phi=\ln y^{2}+K(y-y_{0})^{2} \label{Weylfield}%
\end{equation}
where $K$ is a constant. It is not difficult to verify that the Weyl scalar
field will act as a confining field, trapping massive particles in the
hypersurface $y=y_{0}$. A simple calculation shows that if $K>0$ we are in the
presence of a kind of confinement where particles lying near the hypersurface
$y=y_{0}$\ will oscillate about it, entering and leaving the hypersurface
indefinitely (see (\cite{Dahia}), for details). On the other hand, if $K<0$,
\ the classical confinement is highly unstable. Clearly, the same procedure
can also be used to stabilize the motion of the trapped particles in the case
of Gremm%
\'{}%
s warping function (\ref{warpfunction}). Finally, note that since $\phi
$\ depends only on the extra coordinate $y$, the Riemannian character of the
hypersurfaces $y=const$ is not affected by the presence of the Weyl field.

\section{Final Remarks}

An important class of higher-dimensional models in the braneworld scenario
share the following three properties: a) our spacetime is viewed as
four-dimensional \ Riemannian hypersurface (brane) embedded in a
five-dimensional Riemannian\ manifold (bulk); b) the geometry of the bulk
space is characterized by a warped product space; c) fermionic matter is
confined to the brane by means of an interaction of the fermions with a scalar
field which depends only on the extra dimension. In this article we have
considered the possibility of describing the bulk by a non-Riemannian
geometry, namely, a Weyl manifold. For a class of Weyl fields, the geometry
induced on the brane has a Riemannian structure. However, the confinement and
stability properties of geodesics near the brane may be affected by the Weyl
field. Taking this fact into account we have constructed a classical analogue
of the quantum confinement by considering a Weyl scalar field which also
depends only on the extra coordinate. In a certain way, this Weyl scalar
field, which has a purely geometrical nature, seems to mimic the quantum
scalar field that is responsible for the confinement in field-theoretical
models \cite{Rubakov}.

Throughout this article, we have assumed \ the existence \textit{a priori }of
a Weyl field and have not\ discussed the dynamics of this field and how it
would determine the geometry of the bulk. We leave this subject for a future work.

Finally, as far as the geometrical structure of Weyl inspired
higher-dimensional model is concerned, one would like to look at the embedding
properties of the bulk space. We now know that embedding theorems of
differential geometry are of vital importance for some higher-dimensional
theories of spacetime. This is particularly true in the case of the
induced-matter proposal \cite{Wesson}. Thus, an interesting question is how to
formulate the analogous of the Campbell-Magaard theorem and its extended
versions in the context of a Weyl geometry \cite{Tavakol}. An answer to this
question would, in principle, tell us what kind of Weyl bulk space is
admissible if matter and fields are to be generated from the extra dimensions,
pretty much in the same way as in the case of the (Riemannian) induced matter
proposal and Kaluza-Klein theories.

\bigskip

\section{\bigskip Acknowledgement}

The authors would like to thank CNPq-FAPESQ (PRONEX) for financial support.

\end{document}